\documentstyle[12pt] {article}
\begin {document}
\parindent=15pt
\rightline{{\bf  US-FT/5-96}}
\rightline{{\bf hep-ph/9602230}}
\vspace{1.0cm}
\begin{center}
{\Large{\bf FAST ANTIBARYON 
PRODUCTION IN $pp$\\\vskip 0.3cm COLLISIONS AS A RESULT OF STRING FUSION
}}\\
\vskip 1.5 truecm
{N. Armesto, E. G. Ferreiro, C. Pajares and Yu. M. Shabelski$^\ast$}\\
\vskip 0.3 truecm
{\it Departamento de
F\'{\i}sica de  Part\'{\i}culas,
Universidade de Santiago de
Compostela,\\ 15706-Santiago de Compostela, Spain}\
\end{center}
\vskip 2. truecm
\begin{center}
{\bf ABSTRACT}
\end{center}
\vskip 0.3 truecm

{The inclusion of string fusion in Dual Parton Model results in the
appearance of
diquark-antidiquark pairs in the sea with comparatively large Feynman-$x$.
Such antidiquarks can fragment into fast antibaryons, thus increasing several
times the corresponding yields in high energy $pp$ collisions
for 
$x_F \sim 0.8 \div 0.9$. String fusion also results in an unusual dependence of
inclusive spectra on the multiplicity of secondaries. Some numerical 
estimations are presented.}
\vskip 4.0 truecm
\noindent{$^\ast$Permanent address: Leningrad Nuclear Physics Institute,
Gatchina, Sanct-Petersburg 188350, USSR.}\\
\begin{flushleft}
{\bf US-FT/5-96}\\
{\bf February 1996}
\end{flushleft}

\pagebreak
{\bf 1. INTRODUCTION}
\vskip 0.5 truecm

Models based on pomeron exchange are very popular for the description of
multiparticle production on nucleon and nuclear targets at high energies (see 
for example \cite{KTM,DPM}). In the simplest approach  pomerons are assumed 
not to interact one with each other. In this case the relative content of 
secondaries produced in nucleon-nucleon, nucleon-nucleus and nucleus-nucleus 
collisions should be practically the same because all secondaries are produced 
inside a pomeron (in the sense of the unitarity condition)\footnote{Each cut 
pomeron gives rise to two colour strings streched between partons of the 
projectile and target; the strings decay afterwards into the observed 
secondaries.}. Only very small differences are possible because when a 
different number of pomerons are cut, secondaries are produced at different 
effective energies.

However in high energy heavy ion collisions a relative enhancement of strange 
secondary production in comparison with $pp$ collisions has been experimentally
found \cite{strenh}. The most natural way to explain such effects in the 
framework of pomeron-based models seems to be assuming that pomeron interactions
are more essential in the case of nucleus-nucleus collisions than in the case 
of $pp$ interactions \cite{Paj} (simply because of larger combinatorial 
factors). We discuss below that in the case of the cut of a diagram with 
interacting pomerons colour strings should connect not only valence quarks and 
diquarks and sea quarks, but also with sea diquarks and antidiquarks (coming 
from the fusion of two quark-antiquark strings). These latter fragment 
preferably into baryons and antibaryons so their appearance can change the 
content of secondaries (making possible to explain the strangeness and 
antibaryon enhancement, \cite{Nestor}).

String fusion should exist with a small probability in the case of high energy 
$pp$ collisions and result in antibaryon production with very high Feynman-$x$.
Model estimations predict a yield of such antibaryons several times larger than
in the case of a variant without string fusion. On the other hand the case of 
$pp$ collisions looks more clean in comparison with interactions on nuclear 
targets. So the appearance of such fast antibaryons can be considered as an 
evidence for the existence of this kind of collective effects. For numerical 
estimations in Section 3 we use the Quark-Gluon String Model (QGSM) \cite{KTM} 
which describes quite successfully the spectra of different secondaries 
produced in high energy hadron-nucleon and hadron-nucleus collisions, see for 
example \cite{KTM,12}. This model is very close to the Dual Parton Model 
\cite{DPM}. In particular the spectrum of $\Lambda_s$, to which the 
fragmentation of a valence diquark gives the main contribution\footnote{In the 
case of  high $x_F$ proton production in $pp$ collisions triple-reggeon 
diagrams give the main contribution.} at large $x_F$, is described well enough 
so we can hope that the parameters of antibaryon production by sea-antidiquark 
fragmentation are more or less fixed.  

String fusion also changes the dependence of inclusive spectra on the 
multiplicity. Usually the spectrum of any particle for events with multiplicity
higher than the mean one becomes more and more narrow, due to the division of 
the initial energy between many cut pomerons. However for a particle whose 
spectrum in the large $x_F$ region is determined by string fusion the behaviour
becomes the opposite, i.e., the spectrum gets constant or even broader.

\vskip 0.9 truecm
{\bf 2. INCLUSIVE SPECTRA OF SECONDARY HADRONS IN QGSM} 
\vskip 0.5 truecm
 
In the QGSM high energy hadron-nucleon and hadron-nucleus interactions are 
considered as proceeding via the exchange of one or several pomerons. Each 
pomeron corresponds to a cylindrical diagram, so in the case of one cut pomeron
two showers of secondaries are produced, see Fig. 1. The inclusive spectrum of 
secondaries is determined by the convolution of the diquark and valence and sea
quark distributions $u(x,n)$ in the incident particles with the fragmentation 
functions of quarks and diquarks into secondary hadrons $G(z)$. The diquark and 
quark distribution functions depend on the number $n$ of cut pomerons in the
considered diagram. For a nucleon target the inclusive spectrum of a
secondary hadron $h$ has the form \cite{KTM} 
\begin{equation}
\frac{x_E}{\sigma_{inel}}\  \frac{d\sigma}{dx_F} = 
\sum_{n=1}^{\infty}w_{n}\phi_{n}^{h}(x)\ ,
\end{equation}
where the function $\phi_{n}^{h}(x)$ determines the contribution of the diagram
with $n$ cut pomerons and $w_{n}$ is the probability of this diagram.
We neglect diffractive dissociation, as its effect is known to be
comparatively small in the case of antibaryon production.
 
For $pp$ collisions:
\begin{equation}
\phi_{n}^{h}(x) = f_{qq}^{h}(x_{+},n)f_{q}^{h}(x_{-},n) +
f_{q}^{h}(x_{+},n)f_{qq}^{h}(x_{-},n) +
2(n-1)f_{s}^{h}(x_{+},n)f_{s}^{h}(x_{-},n)\ ,
\end{equation}
\begin{equation}
x_{\pm} = \frac{1}{2}[\sqrt{4m_{T}^{2}/s+x^{2}}\pm{x}]\ ,
\end{equation}
where $f_{qq}$, $f_{q}$ and $f_{s}$ correspond to the contribution to the
inclusive spectrum of 
diquarks, valence and sea quarks respectively. 
They are determined by the convolution
of diquark and quark distributions with fragmentation functions,
\begin{equation}
f_{q}^{h}(x_{+},n) = \int_{x_{+}}^{1} u_{q}(x_{1},n)G_{q}^{h}(x_{+}/x_{1})
\;dx_{1}\ .
\end{equation}
Both diquark and quark distributions and fragmentation functions are 
expressed via their Regge-asymptotics, taking into account the 
conservation laws \cite{KTM,9}.

In the calculations we use quark and diquark distributions 
in the proton of the form \cite{KTM}:
\begin{equation}
u_{uu}(x,n) = C_{uu}\;x^{2.5}(1-x)^{n-1.5}\ , 
\end{equation}
\begin{equation}
u_{ud}(x,n) = C_{ud}\;x^{1.5}(1-x)^{n-1.5}\ ,
\end{equation}
\begin{equation}
u_{u}(x,n) = C_{u}\;x^{-0.5}(1-x)^{n+0.5}\ ,
\end{equation}
\begin{equation}
u_{d}(x,n) = C_{d}\;x^{-0.5}(1-x)^{n+1.5}\ ,
\end{equation}
\begin{equation}
u_{\overline{u}}(x,n) = u_{\overline{d}}(x,n) = C_{\overline{u}}\;x^{-0.5}
[(1+\delta /2)(1-x)^{n+0.5}(1-x/3)-\delta\;(1-x)^{n+1} /2] \; , \; n>1\ ,
\end{equation}
\begin{equation}
u_{s}(x,n) = C_{s}\;x^{-0.5}(1-x)^{n+1} \; , \; n>1\ .
\end{equation}
$\delta =0.2$ is the relative probability to find a strange quark in the 
sea and the factors $C_{i}$ are determined from the normalization condition  
\begin{equation}
\int_{0}^{1} u_{i}(x,n)dx = 1\ .
\end{equation}
Quark and diquark fragmentation functions into secondary hadrons are 
taken from \cite{KTM,9}.

In the case of baryon $B$ production in $pp$ collisions there are two different
contributions \cite{6}. The first one corresponds to the central production of 
a $B\overline{B}$ pair and can be described by the  formulas written above. 
The second contribution is connected with the direct fragmentation of the
initial proton into $B$ with conservation of the string junction. To account 
for this possibility we input into Eq. (2) two additional items 
$f_{qq2}(x_{+},n)$ for $x_F > 0$ and $f_{qq2}(x_{-},n)$ for $x_F < 0$ which 
are not multiplied by $f_{q}(x_{-},n)$ and $f_{q}(x_{+},n)$ respectively. The 
form of $f_{qq2}(x_{+},n)$ and $f_{qq2}(x_{-},n)$ is determined by the 
corresponding fragmentation functions. For example, in the case of $\Lambda_s$ 
production they are  
\begin{equation}
G_{uu2}^{\Lambda_{c}} = a_{02}\;z^{2}(1-z)^{1+\lambda-\alpha_{\varphi}(0)}
\end{equation}
and
\begin{equation}
G_{ud2}^{\Lambda_{c}} = a_{02}\;z^{2}(1-z)^{\lambda-\alpha_{\varphi}(0)}\ ,
\end{equation}
with $\lambda=0.5$ and $\alpha_{\varphi}(0)=0$.

The probability of a process with $n$ cut pomerons is calculated 
in the quasieikonal approximation \cite{KTM,15}:
\begin{equation}
w_{n} = \sigma_{n}/\sum_{n=1}^{\infty}\sigma_{n}  \;  , \;    
\sigma_{n} = \frac{\sigma_{P}}{nz} (1-e^{-z} \sum_{k=0}^{n-1}\frac{z^{k}}{k!
})\ ,
\end{equation}
\begin{equation}
z = \frac{2\;C\;\gamma}{R^{2}+\alpha^{\prime}\xi}e^{\Delta\xi} \; , \; 
\sigma_{P} = 8\pi
\gamma e^{\Delta\xi} \; , \; \xi = \ln(s/1\;GeV^{2}) \ , 
\end{equation}
with
\vskip 0.3 truecm
\[
\Delta = 0.139 \; , \; \alpha^{\prime} = 0.21 \; GeV^{-2}\; , \; 
\gamma_{pp} = 1.77 
\; , \; \gamma_{\pi p} = 1.07 \; , \]
\[
R_{pp}^{2} = 3.18\;GeV^{-2} \; ,
\; R_{\pi p}^{2} = 2.48\;GeV^{-2} \;
, \; C_{pp} = 1.5 \; , \; C_{\pi p} = 1.65\ .\]
\newpage

{\bf 3. STRING FUSION CONTRIBUTION IN QGSM} 
\vskip 0.5 truecm

A possible mechanism of string fusion can be inferred if we consider the 
intermediate
states of a triple-reggeon diagram. Let us consider such a diagram in which a 
triple-reggeon exchange connects sea quarks of two interacting nucleons and let 
all reggeons be cut, Fig. 2a\footnote{Other reggeons which connect
valence quarks are not shown for simplicity.}. One possible inelastic 
intermediate state is shown in Fig. 2b (we present only one of the two produced 
chains, compare to Fig. 1c). If in the upper part 
of the diagrams in Fig. 2 
two pomerons are connected with two sea quarks, in the 
lower part a reggeon will interact with two sea antiquarks because every
chain as a whole has to be white. So the ends of the 
strings (two antiquarks in this case) can fuse and produce some new object which
will fragment into secondary hadrons as a whole. We will call this object a 
sea antidiquark.

In the QGSM an incident fast quark is assumed to fragment into a hadron 
(say, a meson) producing a slower quark which will fragment again and so on, 
until it fuses with a target diquark or antiquark, forming respectively
a baryon or a meson. In the case shown in the Fig. 2b the
same process will occur in the upper part. However a sea diquark will appear at 
a rapidity value approximately equal to the rapidity of the triple-reggeon 
vertex. Then two possibilities appear: 
it can fragment into a baryon $B$ as shown in Fig. 2b or
into mesons until its annihilation with the sea antidiquark. 
It is in the first possibility that we can
obtain a comparatively fast antibaryon in the backward
hemisphere.
Let us note that the lower reggeon in Fig. 2a
corresponds to the pomeron at high energies. This is clear if we compare
the sea diquark-antidiquark interaction in Fig. 2b to the valence  
diquark-antidiquark interaction in the case of high energy $p\overline{p}$
interactions.

The process of fusion of two sea quarks (antiquarks) into one sea diquark 
(antidiquark) becomes possible for $n \geq 3$ in Eq. (1). It seems that the
best place to
see the effects of string fusion is in the spectra of fast antibaryons 
produced in $pp$ collisions because the contribution of all other processes 
is comparatively small. The enhancement of these spectra in 
the large $x_F$ region in the case of string fusion arises from two facts: 
i) the 
$x$-distribution of sea antidiquarks is harder than that of 
sea antiquarks and 
ii) the fragmentation functions are harder.

For the numerical estimations which are, of course, model dependent, we take
the probability of two sea quarks fusing into a diquark equal to $v_0 = 0.02$.
So the probability to find two fused quarks in a diagram with a $n$-pomeron 
exchange is:
\begin{equation}
v_{n} = v_0\; (n-1)\;(2n-3) 
\end{equation}
(the probability that these quarks are both antiquarks is separately
accounted for).
As a result, the total probability for string fusion in
inelastic $pp$ collisions is
\begin{equation}
w_{fus} = \sum_{n=3}^{\infty} v_n \;w_n\ ,
\end{equation}
which is about 1/20 at $\sqrt{s} =$ 39 GeV. This is close to the 
standard contribution of triple-reggeon diagrams at this energy.

For the $x$-distributions of sea antidiquarks we use the simplest assumption: 
it is a convolution of two sea quark distributions,
\begin{equation}
u_{\overline{qq}}(x,n) = \int u_{\overline{q}}(x_1,n) u_{\overline{q}}(x_2,n) 
\;\delta (x-x_1-x_2) \;dx_1 dx_2 
\end{equation}
and we use the standard functions \cite{KTM,9} for their fragmentation into
antibaryons. 

The results for the $x_F$-spectra of $\overline{p}$ and $\overline{\Lambda}_s$ 
produced in $pp$ collisions at different energies are shown in Figs. 3 and 4. 
Model predictions without string fusion are shown by solid lines and with 
string fusion by dashed lines. In all cases the difference in the small $x_F$ 
region is very small and decreasing with energy. The reason is that sea 
antidiquarks have an average $x_F$ value not so small -- about $0.1\div 0.15$, 
so they are far (in rapidity) from the central region. In the large $x_F$ 
region the difference becomes larger and it is about one order of magnitude at 
$x_F = 0.85\div 0.9$ (almost independently on the initial energy).

\vskip 0.9 truecm
{\bf 4. THE SHAPE OF INCLUSIVE SPECTRA IN EVENTS WITH DIFFERENT MULTIPLICITY}
\vskip 0.5 truecm

Let us take a sample of multiparticle production data without low multiplicity 
events, say with $n_{ch} \geq <n_{ch}>$. Usually the spectra of secondaries in 
such a sample become more narrow than that for 
standard events. The reason is that
events corresponding to one-pomeron cut have on the average 
a smaller multiplicity of secondaries
in comparison with multipomeron events, so one-pomeron events are more
frecuently suppressed 
from our high multiplicity sample. Then such a sample is rich in
multipomeron events. But it is in these events where
the $x$-distributions of quarks
and diquarks (ends of strings), Eqs. (5)-(10), are more narrow (simply because
the energy has to be divided between more partons), so these quarks 
and diquarks will generate a more narrow spectrum of secondaries.

However if string fusion is taken into account the behaviour of the spectrum
is more complicated. For secondary mesons or baryons produced in $pp$
collisions we can only expect some small 
quantitative difference because the string
fusion contribution is small practically at all $x_F$. But in the case of
the antibaryon spectrum string fusion dominates the region of
large $x_F$ (Figs. 3 and 4). After substracting the low multiplicity
events the inclusive cross section in this region will change only
slightly\footnote{The 
$x$-distribution of sea antidiquarks is wider than those of sea
antiquarks, as it results from adding the $x$'s of the two parent sea 
antiquarks, Eq. (18).}.
On the other hand it will
decrease more or less significantly in the small $x_F$ region where one-pomeron
processes give about one half of the inclusive cross section. So we can expect
that the spectrum of antibaryons becomes even wider in the set
of high multiplicity events. 

For numerical estimations let us define the variable
\begin{equation}
z = \frac{n_{ch}}{<n_{ch}>}
\end{equation}
and consider the spectra of secondary antibaryons for events with $z \geq z_0$.
$z_0 = 0$ means that we take all events (minimal bias sample), for
$z_0 = 1$ we take only events with multiplicity of charged secondaries larger
than the mean one and so on.

As the differences in the shapes of the are not very large it is better to
consider the ratio of the inclusive cross section in two different points, 
$x_1$ and $x_2$,
\begin{equation}
R\biggl (\frac{x_F = x_1}{x_F = x_2}\biggr ) = 
\frac{[x_E/\sigma_{inel} \ \ d\sigma/dx_F] \vert _{x_F = x_1}}
{[x_E/\sigma_{inel} \ \ d\sigma/dx_F] \vert _{x_F = x_2}}\ ,
\end{equation}
as a function of $z_0$.

The model estimations of these ratios for $x_1 = 0.6$ (where the cross section
is not very small) and $x_2 = 0$ are presented in Fig. 5 for
$\overline{p}$ and $\overline{\Lambda}_s$ produced in $pp$ collisions
at energy $\sqrt{s} = 39$ GeV. It can be seen that in both cases there exists a
large difference in the behaviour of the ratios 
without (solid curves)
and with (dashed curves) string fusion. The experimental
investigation of such ratios could give evidence on the existence of
string fusion.

\vskip 0.9 truecm
{\bf 5. CONCLUSIONS}
\vskip 0.5 truecm

The existence of triple-reggeon interaction has been firmly confirmed by
experiment. The string fusion mechanism is a simple possibility to incorporate 
the contribution of such diagrams into a model of multiparticle production. 
The difference in the content of secondaries between high energy 
nucleon-nucleon and nucleus-nucleus collisions can be explained if one takes 
into accountthe contribution coming from string fusion. However some 
numerically small effect should also exists in the case of high energy $pp$ 
(or $\overline{p}p$) interactions. It can be observed as an enhancement of 
secondary antibaryon production and an unusual dependence of the shapes of 
their inclusive spectra on the multiplicity of
secondaries.

We thank the Direcci\'on General de Pol\'{\i}tica Cient\'{\i}fica and 
the CICYT of Spain for financial support. E.G.F also thanks the Xunta de 
Galicia for for financial support The paper was supported in part by INTAS 
grant 93-0079.

\pagebreak

\newpage

\begin{center}
{\bf Figure captions}
\end{center}
\vskip 0.5 truecm

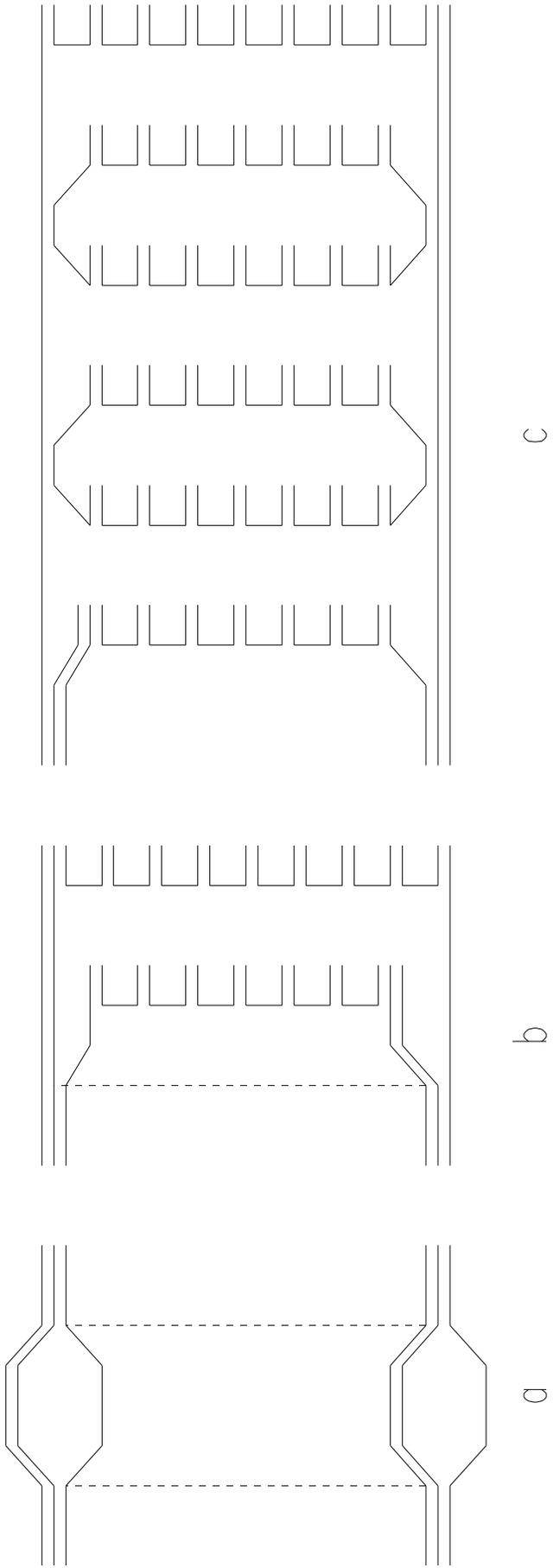
\noindent{\bf Fig. 1.}
Cylindrical diagram which corresponds to the one-pomeron exchange
contribution to elastic $pp$ scattering (a). Its cut which determines the 
contribution to inelastic $pp$ cross section (b). The diagram which
corresponds to the cut of three pomerons (c).

\vskip 0.25 truecm
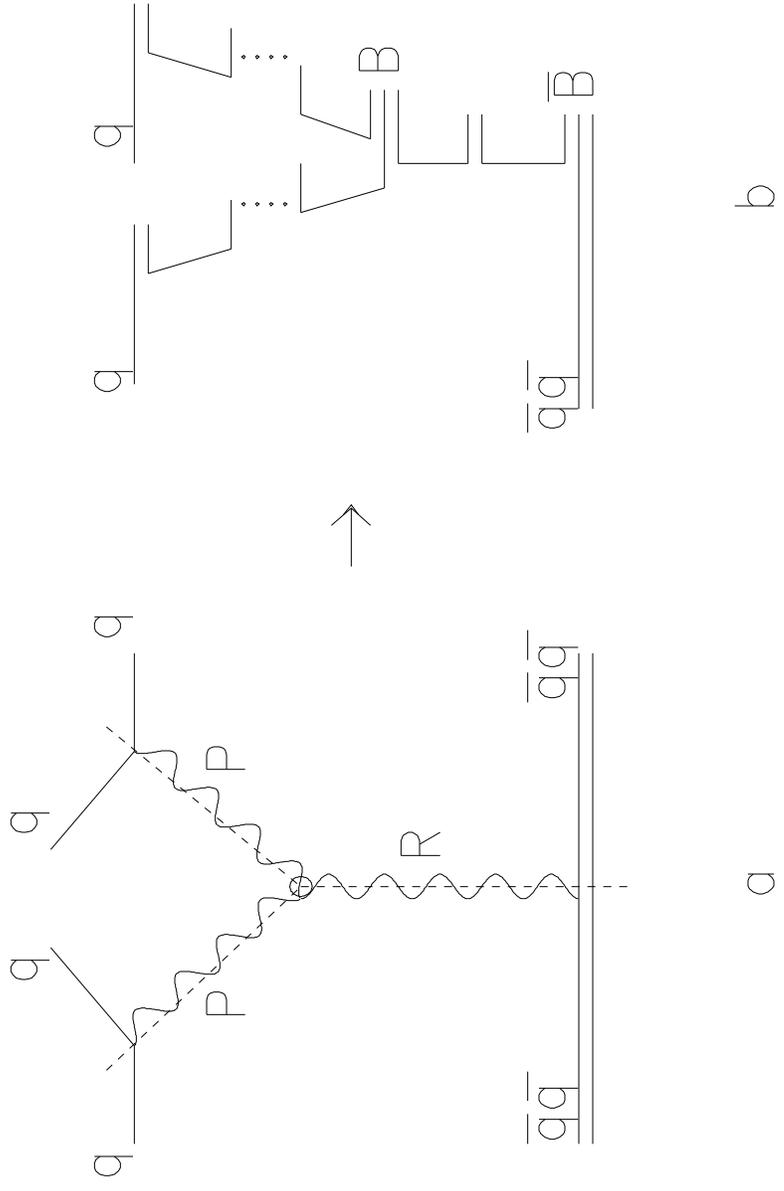
\noindent{\bf Fig. 2.} 
Cut triple-reggeon diagram (a) and the corresponding inelastic
intermediate state (b) with baryon-antibaryon production via string fusion.

\vskip 0.25 truecm
\noindent{\bf Fig. 3.} 
The predicted inclusive spectra of antiprotons in $pp$
collisions at $\sqrt{s}=$ 39 (a), 200 (b) and 1800 (c) GeV 
without (solid curves) and with (dashed curves)
string fusion. 

\vskip 0.25 truecm
\noindent{\bf Fig. 4.} 
The predicted inclusive spectra of $\overline{\Lambda}_s$ in 
$pp$ collisions at $\sqrt{s}=$ 39 (a), 200 (b) and 1800 (c) GeV
without (solid curves) and with (dashed 
curves) string fusion. 

\vskip 0.25 truecm
\noindent{\bf Fig. 5.} The predicted ratios  of $\overline{p}$ (a) and 
$\overline{\Lambda}_s$ (b) 
inclusive spectra in $pp$ collisions at $\sqrt{s} = 39$ 
GeV without (solid curves) and with (dashed curves) string fusion. 


\begin{thebibliography}{99}

\bibitem{KTM} A.B.Kaidalov and K.A.Ter-Martirosyan. Yad.Fiz. 39 (1984) 1545;
40 (1984) 211.

\bibitem{DPM} A.Capella, U.P.Sukhatme, C.-I.Tan and J.Tran Thanh Van. 
Phys.Rep. 236 (1994) 225.\\
K.Werner. Phys.Rep. 232 (1993) 87.

\bibitem{strenh} NA35 Collaboration, presented by D.R\"ohrich and M.Gazdzicki
at the QM'93
Conference, Frankfurt preprints IFK-HENGPG/93-8 (1993),
94-1 (1994) and 93-6 (1993).\\
E.Andersen et al. Phys.Lett. B316 (1993) 603.\\
S.Abatzis  et al. Phys.Lett.
B270 (1991) 123.

\bibitem{Paj}
M.A.Braun and C.Pajares. Phys.Lett.
B287 (1992)
154; Nucl.Phys. B390 (1993) 542; 559.

\bibitem{Nestor}
N.S.Amelin, M.A.Braun and C.Pajares.
Phys.Lett. B306 (1993) 312; Z.
Phys. C-Particles and Fields 63 (1994) 507.\\
N.Armesto, M.A.Braun, E.G.Ferreiro and C.Pajares.
Phys.Lett. B344 (1995) 301.\\
H.Sorge, M.Berenguer, H.St\"ocker and W.Greiner. Phys.Lett.
B289 (1992) 6.

\bibitem{12} Yu.M.Shabelski. Z.Phys. C-Particles and Fields 38 (1988) 569.    

\bibitem{9} Yu.M.Shabelski. Yad.Fiz. 44 (1986) 186.

\bibitem{6} A.B.Kaidalov and O.I.Piskunova. Yad.Fiz. 43 (1986) 1545; Z.
Phys. C-Particles and Fields 30 (1986) 145.
     
\bibitem{15} K.A.Ter-Martirosyan. Phys.Lett. 44B (1973) 377.


\end{thebibliography}
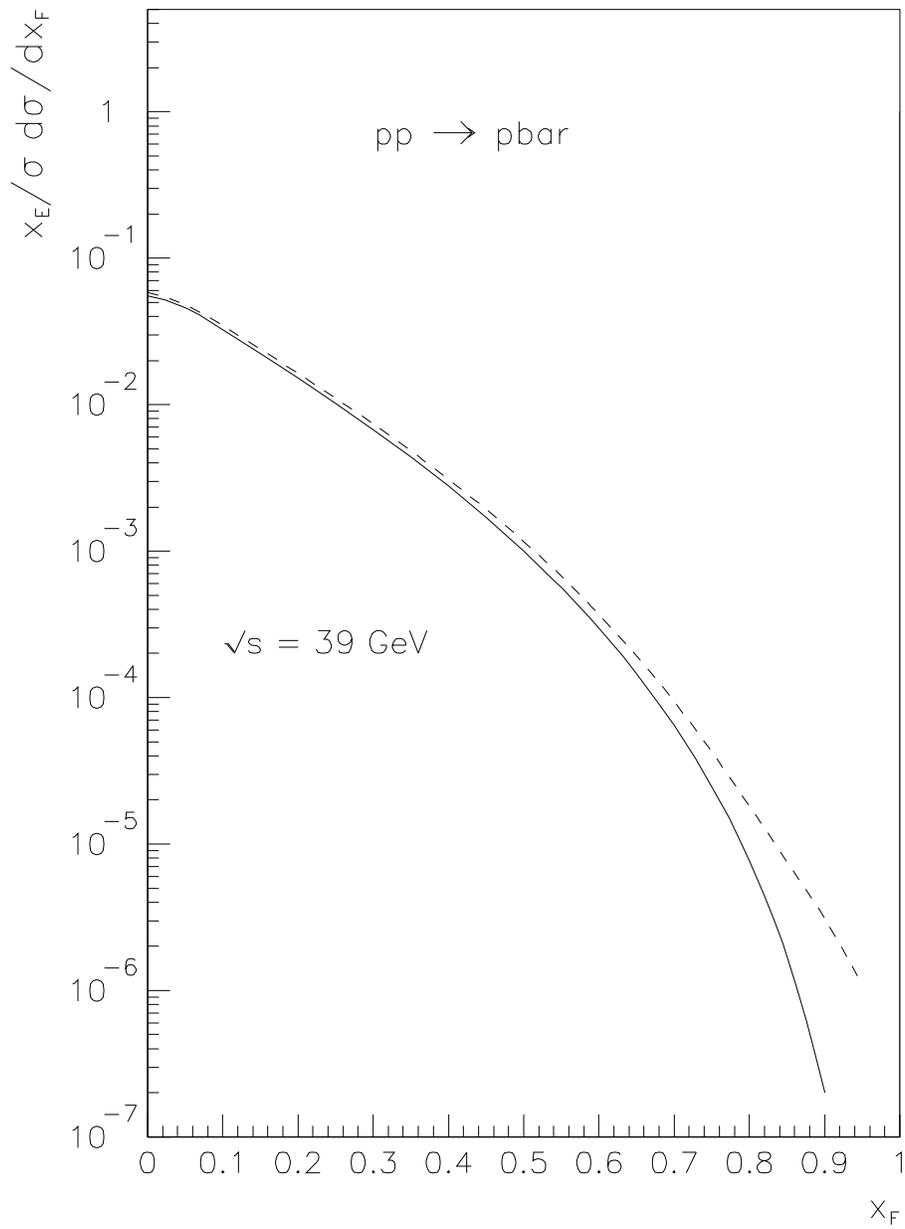
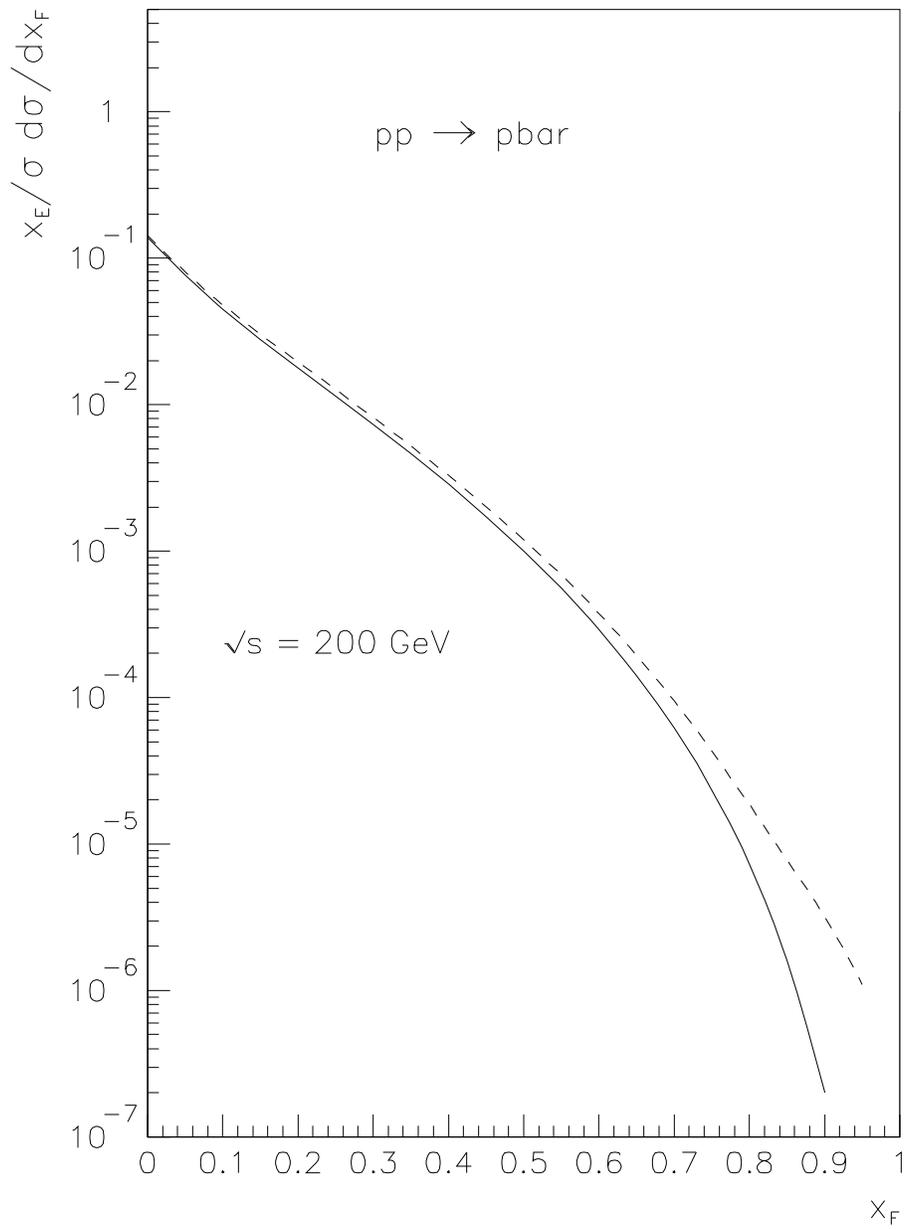
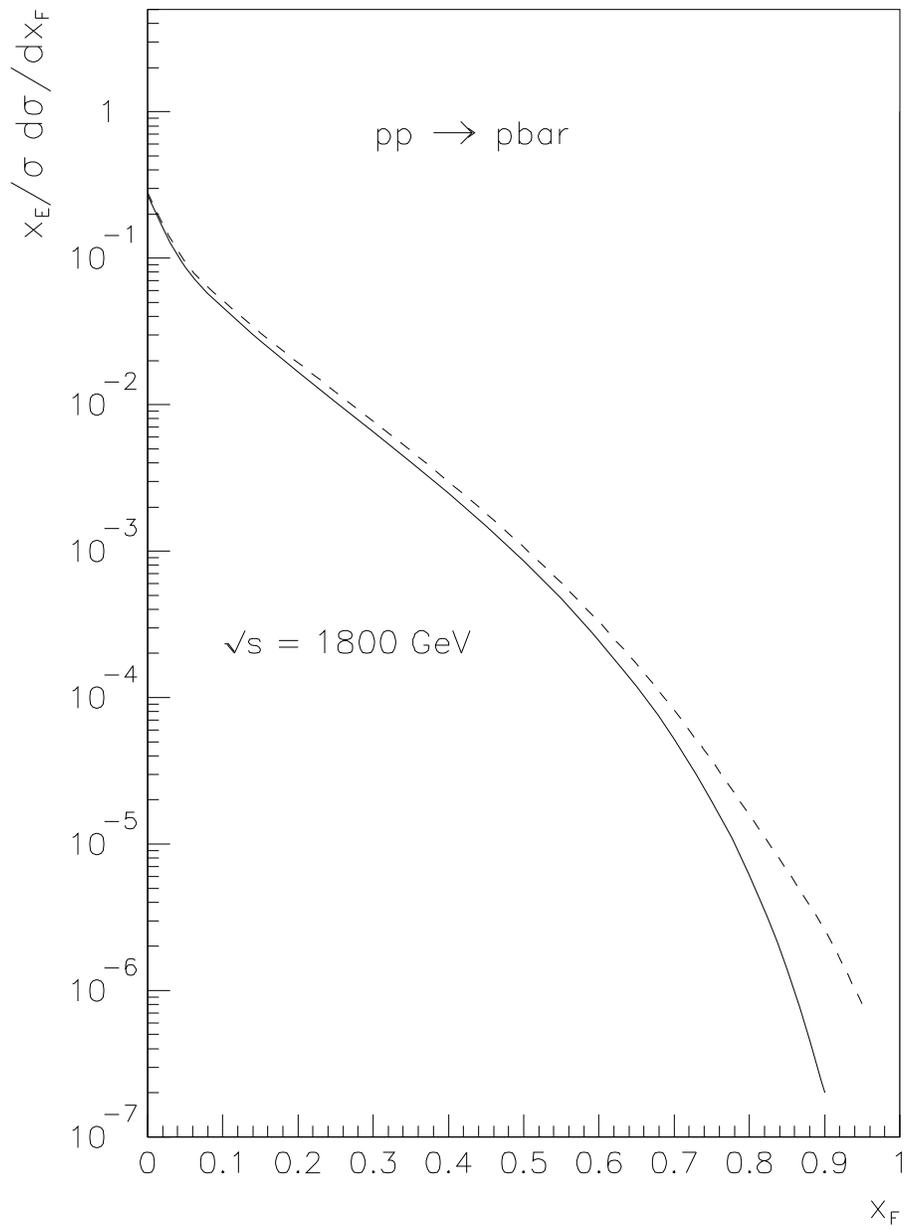
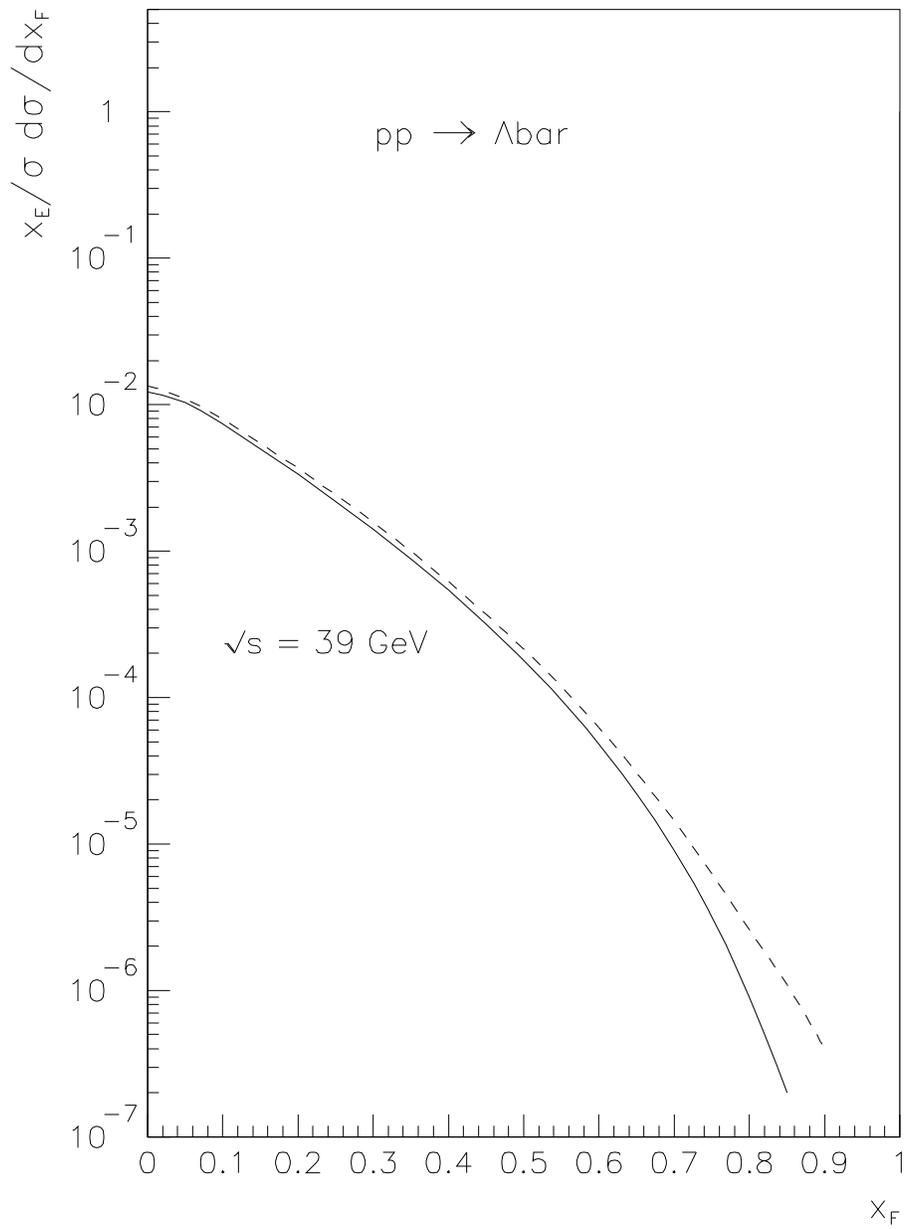
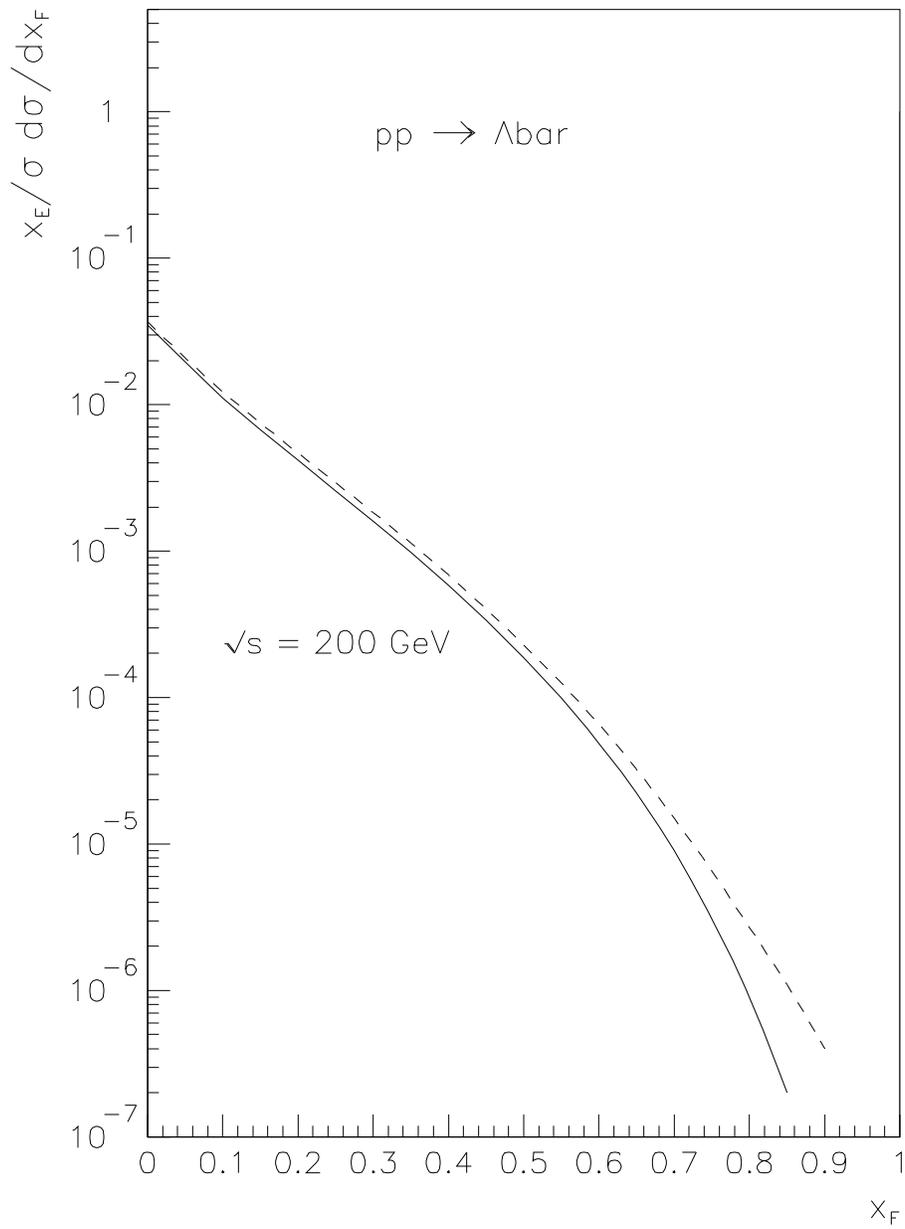
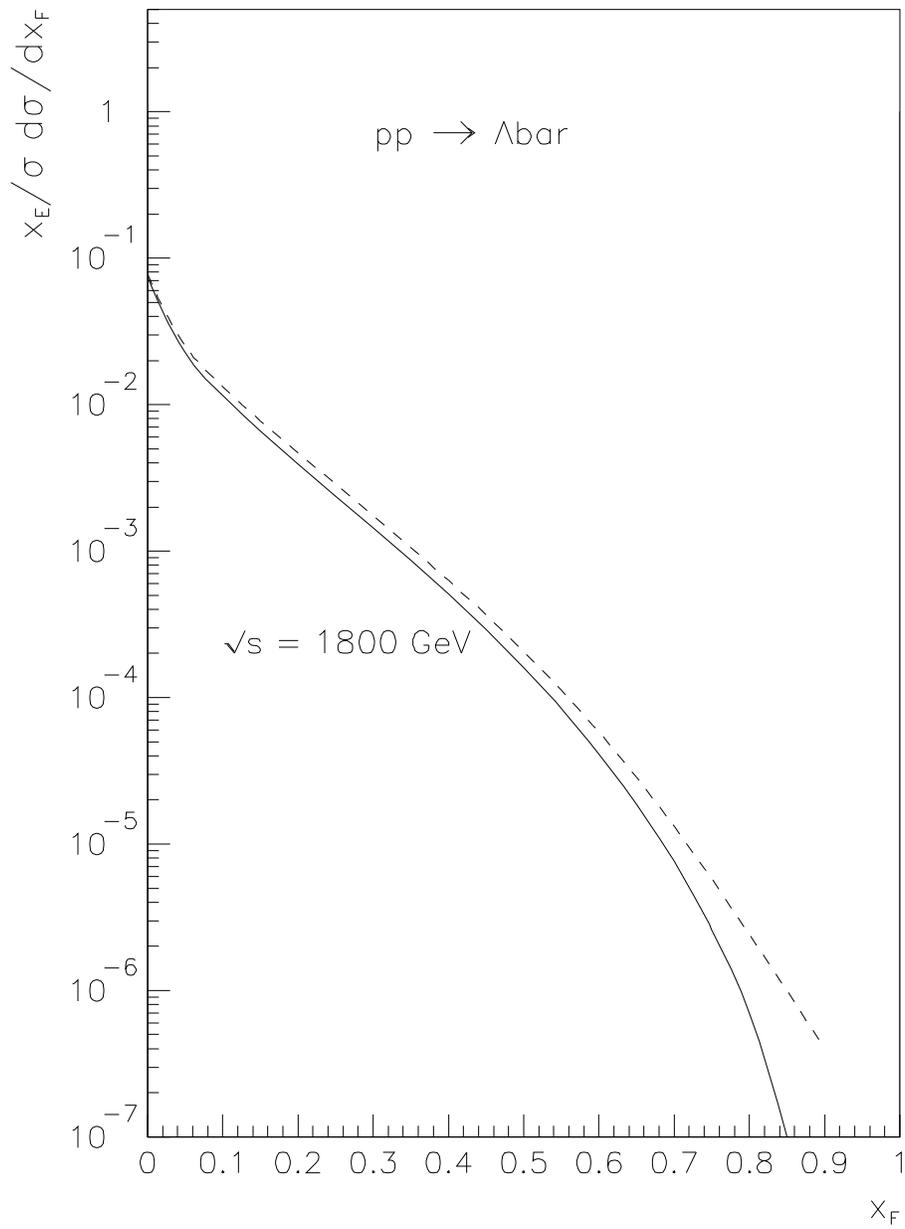
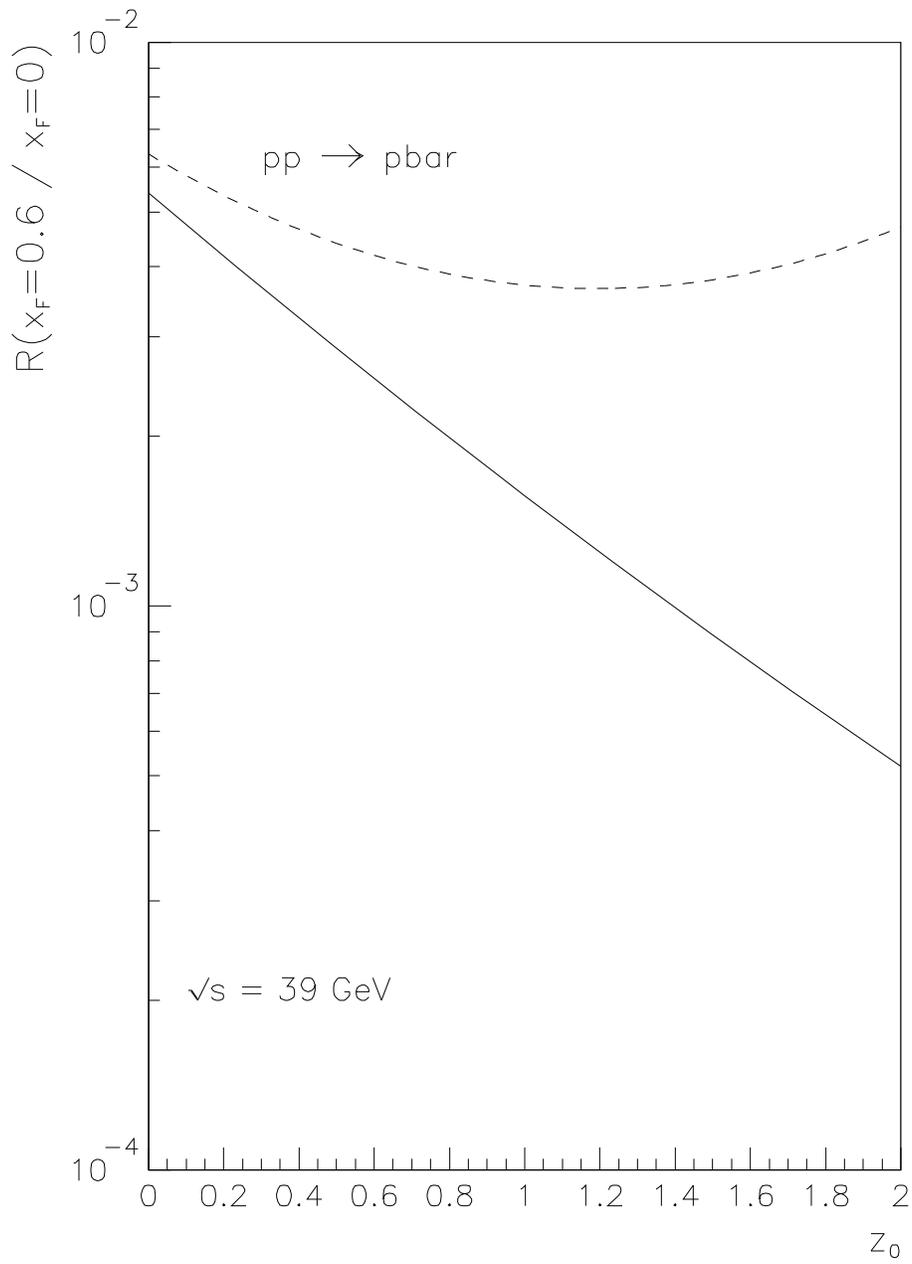
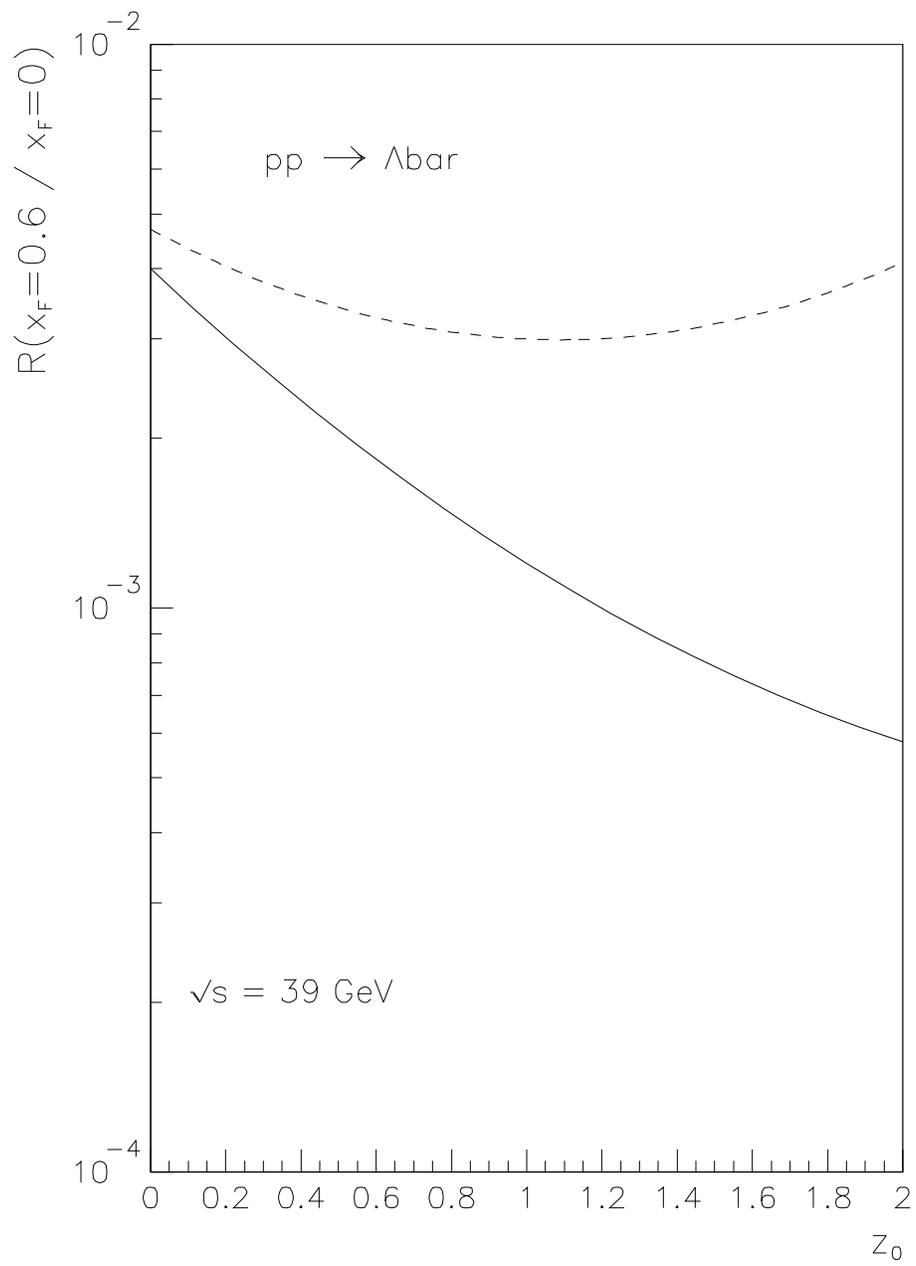
\end{document}